\documentclass[runningheads]{llncs}

\usepackage[T1]{fontenc}
\usepackage{graphicx}
\usepackage{array}
\usepackage{dsfont}
\usepackage{amssymb}
\usepackage{arydshln}
\usepackage{mathtools}
\usepackage{tikz}
\usepackage{multirow}
\usepackage[title]{appendix}
\usepackage[colorlinks=false, hidelinks]{hyperref}

\usetikzlibrary{positioning,decorations.pathreplacing}

\begin{document}
\title{A Data Model and Predicate Logic for Trajectory Data (Extended Version)}

\author{Johann Bornholdt\orcidID{0000-0001-6183-1500} \and
Theodoros Chondrogiannis\orcidID{0000-0002-9623-9133} \and
Michael Grossniklaus\orcidID{0000-0003-1609-2221}}

\authorrunning{J. Bornholdt et al.}

\institute{University of Konstanz, 78457 Konstanz, Germany\\
\email{\{firstname.lastname\}@uni-konstanz.de}
}
\maketitle              
\begin{abstract}
With recent sensor and tracking technology advances, the volume of available trajectory data is steadily increasing. 
Consequently, managing and analyzing trajectory data has seen significant interest from the research community.
The challenges presented by trajectory data arise from their spatio-temporal nature as well as the uncertainty regarding locations between sampled points.
In this paper, we present a data model that treats trajectories as first-class citizens, thus fully capturing their spatio-temporal properties.
We also introduce a predicate logic that enable query processing under different uncertainty assumptions.
Finally, we show that our predicate logic is expressive enough to capture all spatial and temporal relations put forward by previous work.
\end{abstract}

\section{Introduction}\label{sec:introduction}
A growing number of applications ranging from 
rating and publishing personal hiking trips~\cite{chondrogiannis2019inferring}
to studying the migration of animals
require the analysis of trajectory data.
Consequently, the efficient processing of trajectory data has attracted significant interest~\cite{chondrogiannis2022,wang2021survey}.
For example, at the Centre for the Advanced Study of Collective Behaviour\footnote{\url{https://www.exc.uni-konstanz.de/collective-behaviour/}} at the University of Konstanz, the excellence cluster in which the presented research is situated, we are building the so-called \emph{Imaging Hangar}, which enables us to study small animal collectives in a controlled environment using trajectory data obtained from video image analysis~\cite{3d-muppet}.
Depending on the kind of object being tracked, the data recorded together with a trajectory is highly application-specific.
Furthermore, the quality of the trajectory data can vary substantially based on sampling rate and sensor accuracy.

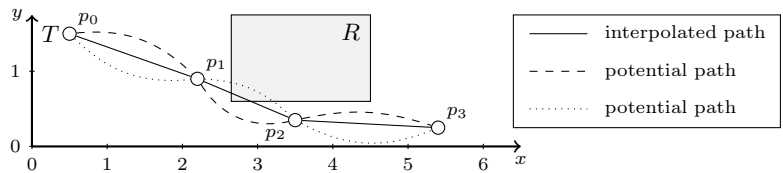
\begin{figure}
    \centering
    \small
    \begin{tikzpicture}
        \tikzset{
            dot/.style={draw, fill=white, circle, inner sep=0pt, minimum size=5},
        }

        \draw[thick,->] (0,0) -- (6.5,0) node[anchor=north] {\scriptsize $x$};
        \foreach \x in {0,1,2,3,4,5,6}
        \draw (\x cm,1pt) -- (\x cm,-1pt) node[anchor=north] {\scriptsize $\x$};
              
        \draw[thick,->] (0,0) -- (0,1.75) node[anchor=east] {\scriptsize $y$};
        \foreach \y in {0,1}
        \draw (1pt, \y cm) -- (-1pt, \y cm) node[anchor=east] {\scriptsize $\y$};

        \draw (6.4, 0.25) rectangle (9.95, 1.75);

        \node at ( 6.5 ,  1.5) (la0) {};
        \node at ( 7.5 ,  1.5) (la1) {};
        \draw (la0) to (la1);
        \node[anchor=west] at (la1) {\scriptsize interpolated path};
        
        \node at ( 6.5 ,  1.0) (lb0) {};
        \node at ( 7.5 ,  1.0) (lb1) {};
        \draw[dashed] (lb0) to (lb1);
        \node[anchor=west] at (lb1) {\scriptsize potential path};

        \node at ( 6.5 ,  0.5) (lc0) {};
        \node at ( 7.5 ,  0.5) (lc1) {};
        \draw[dotted] (lc0) to (lc1);
        \node[anchor=west] at (lc1) {\scriptsize potential path};
        
        \draw[fill = lightgray!20] (2.65,.6) rectangle (4.5,1.75);
        \node[anchor=north west] at (4,1.75) {$R$};
        \node at (.25, 1.5) {$T$};
        
        \node[dot] at ( 0.5 , 1.5  ) (p0) {};
        \node[dot] at ( 2.2 , 0.9  ) (p1) {};
        \node[dot] at ( 3.5 , 0.35 ) (p2) {};
        \node[dot] at ( 5.4 , 0.25 ) (p3) {};

        


        \node[anchor=south west] at (p0) {\scriptsize $p_0$};
        \node[anchor=south west] at (p1) {\scriptsize $p_1$};
        \node[anchor=north east] at (p2) {\scriptsize $p_2$};
        \node[anchor=south west] at (p3) {\scriptsize $p_3$};
        \draw (p0) -- (p1) -- (p2) -- (p3);

        \draw[dashed] (p0) to [bend left=25] (p1);
        \draw[dashed] (p1) to [bend right=35] (p2);
        \draw[dashed] (p2) to [bend left=15] (p3);

        \draw[dotted] (p0) to [bend right=25] (p1);
        \draw[dotted] (p1) to [bend left=20] (p2);
        \draw[dotted] (p2) to [bend right=25] (p3);

    \end{tikzpicture}

    \caption{Example of a trajectory $T$ and a query region $R$.}
    \label{fig:sample_trajectories}
\end{figure}

Trajectory data are uncertain by nature. Specifically, at least two types of uncertainty can be distinguished.
The first type comes from noise in GPS measurements and is inherent to the data source, making it impractical to address it at the system level.
The second type concerns the position of an object between two consecutive trajectory points and is the focus of the work presented in this paper. Due to the discrete sampling rate with which locations are obtained, there is uncertainty as to the exact movement of an object at every point in time.
For example, Figure~\ref{fig:sample_trajectories} shows the trajectory of a bird $T$ and a query region $R$.
Given a straight line between points $p_1$ and $p_2$, $T$ intersects $R$. However, the bird could have actually moved around the corner of $R$, shown as a dashed line.

Existing systems like SECONDO~\cite{guting2005secondo} and MobilityDB~\cite{zimanyi2020mobilitydb} come with two major shortcomings. First, they assume that all necessary trajectories can be collected and stored in a single location. Such a case is not always possible as exchanging trajectory data obtained from different sources has both practical and legal limitations.
Second, to deal with uncertainty, existing approaches either model trajectories using cylinders~\cite{trajcevski2002geometry} and beads~\cite{pfoser1999capturing}, or attempt to process queries by inferring the exact location of the moving object between two recorded locations~\cite{zhan2015range,zheng2011probabilistic}.
However, existing systems do not take the uncertain nature of trajectory data into account. 

To address these shortcomings,   
our aim is to develop a query broker that enables users such as biologists and environmental scientists to query trajectory data from multiple sources through a unified interface.
As a first step, this paper proposes a spatio-temporal predicate logic, including a data model
and operators, to query trajectories from different sources.
In particular, our predicate logic that accommodates uncertainty in trajectory data by supporting different
levels of strictness. The contributions of this paper are as follows.

\begin{itemize}
    \item We introduce a data model for trajectories based on the NF\textsuperscript{2} relational data model. Our model gives equal importance to the spatial and temporal attributes of trajectories while also supporting their application-specific attributes (Section~\ref{sec:trajectories-nf2}).
    \item We define a unified spatio-temporal predicate logic to express selection operations over trajectory data. As a distinguishing feature, our predicate logic supports different levels of strictness to deal with uncertainty in interpreting trajectory data (Section~\ref{sec:selection}).
    \item We demonstrate that our spatio-temporal predicate logic is expressive enough to represent the spatial relations from the DE-9IM standard~\cite{egenhofer1990mathematical} and the temporal relations from Allen's Interval Algebra~\cite{allen1983maintaining}, and we show how our logic handles the uncertainty of trajectory data in a query (Section~\ref{sec:proof-of-concept}).
\end{itemize}
Section~\ref{sec:related-work} provides an overview of existing works.
Concluding remarks and directions for future work are given in Section~\ref{sec:conclusion}.

\section{Related Work}\label{sec:related-work}

In this section, we provide an overview of existing data models, algebras, and systems that have been proposed to store and query trajectory data.

\paragraph*{\bf Data Models and Algebras}

There has been a variety of contributions in the field of data models and algebras for trajectories and moving objects~\cite{bkebel2012olap,bogorny2008spatio,duntgen2010assessing,kucuk2016pg,sadri2004expressing}.
Güting et al.~\cite{guting2000foundation} provide a foundational framework for representing and querying moving objects, which serves as a cornerstone in the trajectory data management domain.
Frihida et al.~\cite{frihida2009modeling} introduce an algebraic spatio-semporal trajectory data type for the representation of trajectory data.
Building on the approach of Frihida et al., Zheni et al.~\cite{zheni2009semantic} introduce a semantic-based model and manipulation language for trajectories. 
A contribution by Ferreira et al.~\cite{ferreira2014algebra} presents an algebra for trajectories by incorporating time series and coverage.
Bakli et al.~\cite{bakli2018spatiotemporal} propose an algebra on operators based on the the Hadoop system.
In contrast to our work, these contributions do not deal with the uncertainty between the sampled points of trajectories in the data model or the algebra.

\paragraph*{\bf Systems}

Several works have contributed to the field of moving object data management~\cite{bakli2019hadooptrajectory,cudre2010trajstore,ding2018ultraman,li2020trajmesa,wang2021survey,wang2018torch}.
Notably, DEDALE~\cite{grumbach1998dedale} is an early system that laid the groundwork for representing and querying moving objects. DEDALE serves as a spatial extension for SQL, lacking a temporal component.
SECONDO~\cite{guting2005secondo} is a research prototype that implements a subset of the foundational framework proposed by Güting et al.~\cite{guting2000foundation}. 
The HERMES trajectory database engine~\cite{pelekis2015hermes} extends the object-relational data model by introducing data types and DDL extensions for managing trajectory data.
MobilityDB~\cite{zimanyi2020mobilitydb} is an extension to PostGIS that supports moving object data providing trajectory-specific data types and functions that implement the DE-9IM relations to a certain degree.
UlTraMan~\cite{ding2018ultraman} extends Apache Spark offering a holistic solution for the entire trajectory pipeline, including range query processing.
Moreover, while most of the aforementioned systems focus primarily on the spatial dimension of trajectory data, time-series database systems~\cite{JPT2017} also support storing and querying trajectory data, focusing primarily on the temporal dimension.
In contrast to our work, most of these systems do not come with a formally defined data model, but instead focus on the technical challenges of trajectory data management. Furthermore, existing systems do not take into account the uncertain aspects of trajectory data.

\section{A Data Model for Trajectories}\label{sec:trajectories-nf2}
Trajectories represent the movement of a moving object. Typically, trajectories are given as a sequence of tuples, that consist of geometric or geographical coordinates accompanied by timestamps.
The timestamp attributes of the tuples of trajectories are strictly ordered and monotonically increasing.
Since most currently available datasets are two-dimensional, we focus on two spatial dimensions.
Definition~\ref{def:trajectory_relation} defines a trajectory in a relational context.

\begin{definition}\label{def:trajectory_relation}
Let trajectory relation $T$ be a relation with schema $sch(T) = {(o, x, y, \tau)}$ that satisfies the following:
\begin{itemize}
    \item
        $\begin{aligned}[t]
            val(T)  &= \lbrace tp | (tp(o), tp(x), tp(y), tp(\tau))\\[-2.5pt]
                    &= \langle (0, x_1, y_1, \tau_1),\dots, (n{-}1, x_{n}, y_{n}, \tau_{n}) \rangle \rbrace, n \in \mathds{N}
        \end{aligned} $
    \item $o$ is the order of each tuple $tp \in T$.
    \item $x$ and $y$ are spatial coordinates (geometrical or geographical).
    \item $\tau$ are timestamps.
    \item $\forall tp_i, tp_j \in T, i \not= j$ it stands that $tp_i(o)> tp_j(o) \Leftrightarrow tp_i(\tau) > tp_j(\tau)$.
\end{itemize}
\end{definition}

In the trajectory relation, we include the \emph{order} column.
While the timestamp could also be used to determine the correct sequence of the tuples in the relation, the order facilitates specific operations, e.g., the retrieval of line segments between consecutive points.
Figure~\ref{fig:sample_trajectory_relational} shows $T$ in a relational table.
It is helpful to store trajectories in relational tables because RDBMS offer a multitude of operators that can be used to run queries on the trajectory data.

\subsection{Trajectory Representation in NF\textsuperscript{2}}

The \emph{Non-First Normal Form} (NF\textsuperscript{2}) data model is an extension of the relational data model. The corresponding NF\textsuperscript{2} algebra allows subexpressions as predicates and enables the access of nested relations. Schek and Scholl~\cite{schek1986relational} provide a detailed description of the algebra. 
In the context of our work, the NF\textsuperscript{2} data model enables the modeling of trajectories as nested relations, thus treating trajectories as first-class citizens.
More specifically, using the NF\textsuperscript{2} data model we store single trajectories $T$ as nested relations of a \emph{trajectories relation} $\mathfrak{T}$, i.e,
\[
\mathfrak{T}(\underline{tid},T(\underline{order},x,y,\tau))
\]
Figure~\ref{fig:sample_nested_trajectory_relational} shows an example of a trajectories relation with a single nested relation representing trajectory $T_0$.

\begin{figure}[t]
    \begin{minipage}[b]{.49\linewidth}
        \centering
        \scriptsize
        \begin{tabular}{|c|c|c|c|}
            \hline
            \textbf{order} & \textbf{x} & \textbf{y} & \textbf{$\tau$} \\
            \hline
            0 & 1.0 & 0.5 & 110 \\
            1 & 2.0 & 1.0 & 120 \\
            2 & 4.0 & 1.5 & 130 \\
            3 & 4.0 & 1.5 & 140 \\
            4 & 3.0 & 0.5 & 150 \\
            \hline
        \end{tabular}
        \caption{The relational representation \\of trajectory $T$.}
        \label{fig:sample_trajectory_relational}
    \end{minipage}
    \begin{minipage}[b]{.49\linewidth}
        \centering
        \scriptsize
        \begin{tabular}{|c|c|}
            \hline
            \textbf{tid} & \textbf{T} \\
            \hline
            \rule{0pt}{9ex} 
            \rule[-8ex]{0pt}{0pt} 
            $T_0$\;\; &
            \begin{tabular}{|c|c|c|c|}%
                \hline
                \textbf{order} & \textbf{x} & \textbf{y} & \textbf{$\tau$} \\
                \hline
                0 & 1.0 & 0.5 & 110 \\
                1 & 2.0 & 1.0 & 120 \\
                2 & 4.0 & 1.5 & 130 \\
                3 & 4.0 & 1.5 & 140 \\
                4 & 3.0 & 0.5 & 150 \\
                \hline
            \end{tabular} \\
            \hline
        \end{tabular}
        \caption{Relational representation $\mathfrak{T}$ \\ of a nested trajectory relation $T_0$.}
        \label{fig:sample_nested_trajectory_relational}
    \end{minipage}
\end{figure}

\subsection{Data Point and Trajectory Properties}

In order to represent additional properties of a trajectory, we can add a relation with properties referencing the trajectory relation.
For properties that apply to entire trajectories, we add a column to the properties relation for each property.
Trajectory properties can be added with two different scopes:
\begin{enumerate}
    \item \emph{trajectory properties}, for properties on entire trajectories.
    \item \emph{point properties}, for properties on trajectory points.
\end{enumerate}

Figure~\ref{fig:trajectory-properties} shows a trajectory property relation with one example column for each property type.
A \emph{trajectory property} is shown in the \emph{species} column.
It contains the type of animal, that was tracked for this trajectory.
As a trajectory property, its value applies to the entire trajectory.
A \emph{point property} enables the storage of a specific property associated with a single point of the trajectory.
The column storing point properties contains a nested relation with the \emph{order} of the point and the corresponding property.
It is important for the consistency of the data model, to relate it to the order instead of the timestamp, to enforce that properties always correspond to specific points of the trajectory.
Point properties can be helpful for properties that only apply to a few points.
For example, in the properties relation shown in Figure~\ref{fig:trajectory-properties}, the \emph{movement type} column contains information about how the bird is moving, e.g., flying or walking.

\subsection{Segment Property Uncertainty}\label{sec:segment-property-uncertainty}

An inherent problem when representing the movement of an object using trajectories is that observations can only be captured at distinct timestamps.
However, the actual movement of an object is continuous. As such, when processing a query, we cannot assume that any point properties also apply to locations on the segments between two consecutive points.
Figure~\ref{fig:segment-properties} shows a segment interpretation of the point properties in Figure~\ref{fig:trajectory-properties}.
Between points $0$ and $1$, i.e., between timestamps $110$ and $120$, the movement type property has the value ``walking'' and between points $2$ and $4$, i.e., timestamps $130$ and $150$, the value ``flying''.
Hence, whether the value ``walking,'' applies for timestamp $115$ depends on the semantics of the point property. Even so, no assumption can be made about the value of the property, e.g., for the timestamp $125$, i.e., it is uncertain at which point in time the value of the property changes. 
Due to this uncertainty, we do not consider this type of property in the data model. Instead, we consider such properties only in the context of a query and we propose a strictness option in Section~\ref{sec:selection-uncertainty} to deal with the uncertainty.

 \begin{figure}[t]
    \begin{minipage}[b]{.49\linewidth}
        \centering
        \scriptsize
        \begin{tabular}{|c|c|c|c|}
            \hline
            \textbf{tid} & \textbf{species} & \textbf{movement type} \\
            \hline
            \rule{0pt}{9ex} 
            \rule[-8ex]{0pt}{0pt} 
            $T_0$\;\; &
            goose &
            \begin{tabular}{|c|c|}%
                \hline
                \textbf{order} & \textbf{movement type} \\
                \hline
                0 & walking \\
                1 & walking \\
                2 & flying \\
                3 & flying \\
                4 & flying \\
                \hline
            \end{tabular} \\
            \hline
        \end{tabular}
        \caption{Property relation for trajectories (species) and points (movement type).}
        \label{fig:trajectory-properties}
    \end{minipage}
    \begin{minipage}[b]{.49\linewidth}
        \centering
        \scriptsize
        \begin{tabular}{|c|c|c|}%
            \hline
            \textbf{begin} & \textbf{end} & \textbf{movement type} \\
            \hline
            0 & 1 & walking \\
            2 & 4 & flying \\
            \hline
        \end{tabular}
        \caption{Query result when interpreting consecutive point properties continuously.}
        \label{fig:segment-properties}
    \end{minipage}
\end{figure}

\section{Spatio-Temporal Predicate Logic}\label{sec:selection}
The selection operator is an essential query operator in database systems that filters tuples based on a given predicate.
In this section, we introduce the spatio-temporal selection $\sigma^{ST}$ operator that performs a range selection over a set of trajectories on the spatial and temporal dimensions. The operator applies a spatio-temporal predicate $P$ on a relation of trajectories $\mathfrak{T}$.
The result is a subset of the trajectories in $\mathfrak{T}$ which satisfy $P$.

\begin{definition}
Given a trajectories relation $\mathfrak{T}$, and a spatio-temporal predicate $P$, the spatio-temporal trajectory selection $\sigma^{ST}$ returns a relation that contains every tuple $tp$ in $\mathfrak{T}$ for which the trajectory relation $tp.T$ satisfies $P$, i.e.,
\begin{gather*}
sch(\sigma^{ST}_P(\mathfrak{T})) = sch(\mathfrak{T})\\
val(\sigma^{ST}_P(\mathfrak{T})) = \lbrace tp : tp \in \mathfrak{T} \wedge P \text{ holds for } tp.T \rbrace
\end{gather*}
\end{definition}

To express spatio-temporal predicates, we have designed a predicate notation that works equally for the spatial and temporal dimensions.
Our notation can be used to define specific conditions on the points of trajectories.
When using a predicate in a selection operator, all of these conditions must be satisfied for a trajectory to be in the result set. Additionally, multiple predicates can be combined in conjunctive normal form.
For example, let a query $Q$ be: ``find all trajectories in an region $R$ during a time interval $I$''.
Figures~\ref{fig:spatial-seleciton} and~\ref{fig:temporal-seleciton} show a visualization of the spatial and temporal portion of the query, respectively.
In the following subsections, we introduce our predicate logic and show how it can independently solve the spatial and temporal parts of $Q$.

\subsection{Spatial Predicates}\label{sec:spatial-predicates}

The spatial part of a predicate is used to express a two-dimensional range query.
The predicates are applied to trajectories on a point level.
The following objects can be used in spatial predicates:

\begin{itemize}
    \item $T$: The trajectory relation that contains all points in the trajectory,
    \item $R$: A geometric object,
    \item $\{p_f, p_l\} \in T$: The first and last point of the trajectory,
    \item $T_{\widehat{fl}} = T \setminus\{p_f, p_l\}$: All points of the trajectory except the first and the last.
\end{itemize}

\begin{figure}[t]
    \begin{minipage}[b]{.49\linewidth}
        \centering
        \small
        \begin{tikzpicture}
            \tikzset{
                dot/.style={draw, fill=white, circle, inner sep=0pt, minimum size=5},
            }
            
            \draw[thick,->] (0,0) -- (5.25,0) node[anchor=north] {$x$};
            \foreach \x in {0,1,2,3,4,5}
              \draw (\x cm,1pt) -- (\x cm,-1pt) node[anchor=north] {$\x$};
              
            \draw[thick,->] (0,0) -- (0,2.25) node[anchor=east] {$y$};
            \foreach \y in {0,1,2}
              \draw (1pt, \y cm) -- (-1pt, \y cm) node[anchor=east] {$\y$};
            
            \draw[draw=black, fill = lightgray!20] (1.5,.5) rectangle ++(3,1);
            \node[anchor=north west] at (1.5cm, 1.5cm) {$R$};
            
            \node[anchor=east] at (.75,1.5) {$T_a$};
            \node[dot] at (1, 1.5) (pa1) {};
            \node[anchor=north] at (pa1) {\scriptsize $a_1$};
            \node[dot] at (2, 1.75) (pa2) {};
            \node[anchor=north] at (pa2) {\scriptsize $a_2$};
            \node[dot] at (3, 1.75) (pa3) {};
            \node[anchor=north] at (pa3) {\scriptsize $a_3$};
            \node[dot] at (4, 1.4) (pa4) {};
            \node[anchor=north] at (pa4) {\scriptsize $a_4$};
            \node[dot] at (3, 0.8) (pa5) {};
            \node[anchor=north] at (pa5) {\scriptsize $a_5$};
            \draw (pa1) -- (pa2) -- (pa3) -- (pa4) -- (pa5);
            
            \node[anchor=east] at (.75,1) {$T_b$};            
            \node[dot] at (1, 1) (pb1) {};
            \node[anchor=north] at (pb1) {\scriptsize $b_1$};
            \node[dot] at (2, 1.125) (pb2) {};
            \node[anchor=north] at (pb2) {\scriptsize $b_2$};
            \node[dot] at (3, 1.25) (pb3) {};
            \node[anchor=north] at (pb3) {\scriptsize $b_3$};
            \node[dot] at (4, 1) (pb4) {};
            \node[anchor=north] at (pb4) {\scriptsize $b_4$};
            \node[dot] at (5, 1.25) (pb5) {};
            \node[anchor=north] at (pb5) {\scriptsize $b_5$};
            \draw (pb1) -- (pb2) -- (pb3) -- (pb4) -- (pb5);
            
            \node[anchor=east] at (.75,.5) {$T_c$};
            \node[dot] at (1, 0.5) (pc1) {};
            \node[anchor=north] at (pc1) {\scriptsize $c_1$};
            \node[dot] at (2, .3) (pc2) {};
            \node[anchor=north] at (pc2) {\scriptsize $c_2$};
            \node[dot] at (3, .3) (pc3) {};
            \node[anchor=north] at (pc3) {\scriptsize $c_3$};
            \node[dot] at (4, .3) (pc4) {};
            \node[anchor=north] at (pc4) {\scriptsize $c_4$};
            \node[dot] at (5, .4) (pc5) {};
            \node[anchor=north] at (pc5) {\scriptsize $c_5$};
            \draw (pc1) -- (pc2) -- (pc3) -- (pc4) -- (pc5);
        \end{tikzpicture}
        \caption{Spatial component of $Q$}
        \label{fig:spatial-seleciton}
    \end{minipage}
    \begin{minipage}[b]{.49\linewidth}
        \centering
        \small
        \begin{tikzpicture}
            \tikzset{
                dot/.style={draw, fill=white, circle, inner sep=0pt, minimum size=5},
            }
            
            \fill[fill = lightgray!20] (0,0) rectangle ++(2,2);
            \node[anchor=north west] at (0cm, 2cm) {$I$};
            \draw (0,0) -- (0,2);
            \draw (2,0) -- (2,2);
                
            \draw[thick,->] (-0.5,0) -- (3.5,0) node[anchor=north] {$t$};
            \draw (0 cm, 1pt) -- (0 cm, -1pt) node[anchor=north] {$100$};
            \draw (0.5 cm, 1pt) -- (0.5 cm, -1pt);
            \draw (1 cm, 1pt) -- (1 cm, -1pt) node[anchor=north] {$120$};
            \draw (1.5 cm, 1pt) -- (1.5 cm, -1pt);
            \draw (2 cm, 1pt) -- (2 cm, -1pt) node[anchor=north] {$140$};
            \draw (2.5 cm, 1pt) -- (2.5 cm, -1pt);
            \draw (3 cm, 1pt) -- (3 cm, -1pt) node[anchor=north] {$160$};
            
            \node[anchor=east] at (-.25,1.5) {$T_a$};
            \node[dot] at (.5, 1.5) (pa1) {};
            \node[anchor=north] at (pa1) {\small $a_1$};
            \node[dot] at (1, 1.5) (pa2) {};
            \node[anchor=north] at (pa2) {\small $a_2$};
            \node[dot] at (1.5, 1.5) (pa3) {};
            \node[anchor=north] at (pa3) {\small $a_3$};
            \node[dot] at (2, 1.5) (pa4) {};
            \node[anchor=north] at (pa4) {\small $a_4$};
            \node[dot] at (2.5, 1.5) (pa5) {};
            \node[anchor=north] at (pa5) {\small $a_5$};
            \draw (pa1) -- (pa2) -- (pa3) -- (pa4) -- (pa5);
            
            \node[anchor=east] at (-.25,1) {$T_b$};
            \node[dot] at (0, 1) (pb1) {};
            \node[anchor=north] at (pb1) {\small $b_1$};
            \node[dot] at (.5, 1) (pb2) {};
            \node[anchor=north] at (pb2) {\small $b_2$};
            \node[dot] at (1, 1) (pb3) {};
            \node[anchor=north] at (pb3) {\small $b_3$};
            \node[dot] at (1.5, 1) (pb4) {};
            \node[anchor=north] at (pb4) {\small $b_4$};
            \node[dot] at (2, 1) (pb5) {};
            \node[anchor=north] at (pb5) {\small $b_5$};
            \draw (pb1) -- (pb2) -- (pb3) -- (pb4) -- (pb5);
            
            \node[anchor=east] at (-.25,.5) {$T_c$};            
            \node[dot] at (1, .5) (pc1) {};
            \node[anchor=north] at (pc1) {\small $c_1$};
            \node[dot] at (1.5, .5) (pc2) {};
            \node[anchor=north] at (pc2) {\small $c_2$};
            \node[dot] at (2, .5) (pc3) {};
            \node[anchor=north] at (pc3) {\small $c_3$};
            \node[dot] at (2.5, .5) (pc4) {};
            \node[anchor=north] at (pc4) {\small $c_4$};
            \node[dot] at (3, .5) (pc5) {};
            \node[anchor=north] at (pc5) {\small $c_5$};
            \draw (pc1) -- (pc2) -- (pc3) -- (pc4) -- (pc5);

        \end{tikzpicture}
        \caption{Temporal component of $Q$}
        \label{fig:temporal-seleciton}
    \end{minipage}
\end{figure}

Without loss of generality, we consider the case of spatial range queries, where $R$ represents a region.
While on a logical level, $R$ can be any two-dimensional shape, the nature of the physical implementation of the predicates can affect the supported shapes of $R$.
Furthermore, the calculation of the predicates changes depending on whether $R$ is, for example, a rectangle or a polygon.
Since the most common spatial range queries are bounding boxes, we focus on rectangle-shaped regions in the upcoming examples.

The relationship between points of a trajectory $T$ and $R$ can be expressed with the following operators: \emph{contained} ($\sqsubseteq$), \emph{properly contained} ($\sqsubset$), and \emph{not contained} ($\not\sqsubset$).
The \emph{contained} operator asserts that one or several points must be inside $R$ or on the border of $R$.
It can be used to model the spatial portion of query $Q$ shown in Figure~\ref{fig:spatial-seleciton} by asserting that at least one point in $T$ must be contained in $R$ with the predicate:
$$\exists p \in T: p \sqsubseteq R$$
In order to ensure that the entire trajectory $T$ is contained in (or on the border of) $R$, a predicate can be written as:
$$\forall p \in T: p \sqsubseteq R$$
The \emph{properly contained} operator functions similarly to \emph{contained} but excludes points that lie on the border of $R$.
In addition to expressing that points are properly contained in $R$, \emph{properly contained} can also be used in conjunction with \emph{contained} to express that points are on the border of $R$:
$$\forall p \in T: p \sqsubseteq R \wedge \neg( p \sqsubset R )$$
Lastly, the \emph{not contained} operator enables the construction of predicates for points that are disjoint from $R$:
$$\forall p \in T : p \not\sqsubset R$$

For some complex spatial relations, it is necessary to address a trajectory's first or last parts specifically.
To achieve this, we introduce two unique points, $p_f$ and $p_l$, denoting the trajectory's first and last points $T$, respectively.
These special points can be used, for example, to express that the beginning of $T$ is contained in $R$, but the end of $T$ is not:
$$p_f \sqsubset R \wedge p_l \not \sqsubset R$$

To enhance the conciseness of the formulas, we utilize the subset $T_{\widehat{fl}}$ of $T$, which includes all points except $p_f$ and $p_l$.
Using $T_{\widehat{fl}}$, a predicate for a trajectory that starts and ends inside $R$, but has points outside of $R$ is expressed as:
$$p_f, p_l \sqsubset R \wedge \exists p \in T_{\widehat{fl}}: p \not \sqsubset R$$

\subsection{Temporal Predicates}\label{sec:temporal-predicates}

The operators introduced in Section~\ref{sec:spatial-predicates} ($\sqsubset$, $\sqsubseteq$, $\not\sqsubset$) for spatial predicates can also be used to express the relationship between trajectories points and a time interval $I$.
In contrast to the two-dimensional region $R$, $I$ is one-dimensional, which makes the start and endpoints of $I$ its \emph{border}.
Figure~\ref{fig:temporal-seleciton} shows the temporal part of query $Q$, where the border points of $I$ are at timestamps 1 and 4.
The temporal part of $Q$ can be expressed with the introduced operators:
$$\exists p \in T: p \sqsubseteq I$$

To effectively express temporal relations, we introduce two additional operators: \emph{p is before I} ($p < I$), and \emph{p is after I} ($p > I$).
The operator $p < I$ expresses that point p is earlier on the time axis than the start of the interval $I$, for $p > I$, $p$ is after the end point of $I$.
Therefore, both operators express that $p$ is outside of $I$.
For example, a predicate for a trajectory that completely overlaps interval $I$ can be expressed as:
$$ p_f < I \wedge p_l > I$$

\subsection{Combining Spatial and Temporal Predicates}\label{sec:spatio-temporal_predicates}

One strength of our predicate logic is its ability to combine spatial and temporal predicates seamlessly.
Users can express complex spatio-temporal queries involving the geometric characteristics of trajectories and their temporal evolution.

When looking at the example query $q$, we can now express the spatial and temporal parts of the query.
However, suppose we express both independently from each other. In that case, we can see in Figures~\ref{fig:spatial-seleciton}~and~\ref{fig:temporal-seleciton} that $T_a$ intersects both spatial region $R$, and interval $I$, but not in the same points.
To properly express $Q$, we need to define a predicate where a single point $p$ of trajectory $T$ is both in $R$ and $I$:
$$\exists p \in T: p \sqsubseteq R \wedge p \sqsubseteq I$$
In our predicate logic, spatio-temporal predicates can be expressed easily because the same operators can be used on both the spatial and the temporal dimensions.

\subsection{Selection Uncertainty}\label{sec:selection-uncertainty}

As discussed in Section~\ref{sec:segment-property-uncertainty}, the segments between individual points are not known in trajectories.
For the spatio-temporal selection, these unknown segments pose a problem because the predicate operators introduced in Sections~\ref{sec:spatial-predicates} and~\ref{sec:temporal-predicates} are applied on sets of points.
However, when using the concrete points of trajectories, we are not examining the segment between points.

Figure~\ref{fig:sample_trajectories} shows a bird's trajectory $T$ and a query region $R$.
To check whether a trajectory $T$ intersects with a region $R$, assume a predicate $Q_1 = \exists p \in T: p \sqsubset R$, which checks if $T$ has a point which is contained in $R$.
On a point-by-point evaluation, $T$ does not satisfy $q_1$ because all points of $T$ lie outside of $R$, even though we can see in Figure~\ref{fig:sample_trajectories} that the interpolation of $T$ intersects $R$.
The same uncertainty also exists in the inverse query, when checking whether $T$ does not intersect $R$ with the predicate $Q_2 = \forall p \in T: p \not\sqsubset R$.

To tackle this uncertainty in the spatio-temporal predicates, we propose a \emph{strictness parameter}, added to the predicates, to define how ambiguous trajectories will be treated for each predicate. We identify three degrees of strictness:

\begin{itemize}
    \item The \emph{strict} evaluation of predicates considers only the points of the trajectory. In the example above, the strict evaluation of \emph{$Q_1$} does not match $T$, while \emph{$Q_2$} does match $T$.
    \item The \emph{relaxed} evaluation of predicates assumes that there are infinitely many intermediate points on the straight line segments between pairs of consecutive points in trajectories, thereby assuring that all intersections between $T$ and $R$ are considered. In the example above, the relaxed evaluation of \emph{$Q_1$} matches $T$, while \emph{$Q_2$} does not match $T$.
    \item The \emph{approximated} evaluation allows users to inject custom behavior into the predicates for cases where the \emph{strict} and \emph{relaxed} evaluations are not sufficient, e.g., for restricted trajectories.
    %
\end{itemize}

With the strictness parameter, the user can define in a query whether the trajectory should be evaluated as a set of points (strict), as a continuous movement (relaxed), or by using some user-define assumption about the movement of the object (approximated). As investigating multiple assumptions for the \emph{approximated} evaluation is out of the scope of our paper, we focus on strict and relaxed evaluation. This strictness parameter applies for spatial predicates as well as for the segment properties described in Section~\ref{sec:segment-property-uncertainty}.

\section{Proof of Concept}\label{sec:proof-of-concept}
In this section, we demonstrate the completeness of our predicate operators, introduced in Section~\ref{sec:selection} for spatio-temporal range queries by establishing their equivalence and compatibility with established relationship models.

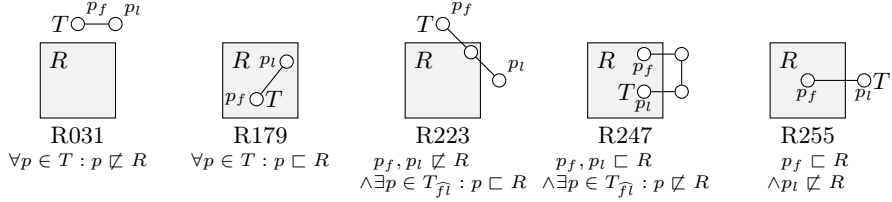
\begin{figure}[t]
    \begin{minipage}[b]{.19\linewidth}
        \centering
        \small
        \begin{tikzpicture}
            \tikzset{
                dot/.style={draw, fill=white, circle, inner sep=0pt, minimum size=5},
            }
            \useasboundingbox (-0.5,-0.25) rectangle (1.5,1.5);
            \draw[fill = lightgray!20] (0,0) rectangle (1,1);
            \node[anchor=north west] at (0,1) {$R$};
               
            \node[dot] at (0.5, 1.25) (pf) {};
            \node[anchor=east] at (pf) {$T$};
            \node[anchor=south west] at (pf) {\scriptsize $p_f$};
            \node[dot] at (1, 1.25) (pl) {};
            \node[anchor=south west] at (pl) {\scriptsize $p_l$};
            \draw (pf) -- (pl);
            
            \node[anchor=north] at (.5,0) {R031};
        \end{tikzpicture}
    \end{minipage}
    \begin{minipage}[b]{.19\linewidth}
        \centering
        \small
        \begin{tikzpicture}
            \tikzset{
                dot/.style={draw, fill=white, circle, inner sep=0pt, minimum size=5},
            }
            \useasboundingbox (-0.5,-0.25) rectangle (1.5,1.5);
            \draw[fill = lightgray!20] (0,0) rectangle (1,1);
            \node[anchor=north west] at (0,1) {$R$};
               
            \node[dot] at (0.45, 0.25) (pf) {};
            \node[anchor=west] at (pf) {$T$};
            \node[anchor=east] at (pf) {\scriptsize $p_f$};
            \node[dot] at (0.85, 0.75) (pl) {};
            \node[anchor=east] at (pl) {\scriptsize $p_l$};
            \draw (pf) -- (pl);
            
            \node[anchor=north] at (.5,0) {R179};
        \end{tikzpicture}
    \end{minipage}
    \begin{minipage}[b]{.19\linewidth}
        \centering
        \small
        \begin{tikzpicture}
            \tikzset{
                dot/.style={draw, fill=white, circle, inner sep=0pt, minimum size=5},
            }
            \useasboundingbox (-0.5,-0.25) rectangle (1.5,1.5);
            \draw[fill = lightgray!20] (0,0) rectangle (1,1);
            \node[anchor=north west] at (0,1) {$R$};
               
            \node[dot] at (0.5, 1.25) (pf) {};
            \node[anchor=east] at (pf) {$T$};
            \node[anchor=south west] at (pf) {\scriptsize $p_f$};
            \node[dot] at (0.875, 0.875) (p) {};
            \node[dot] at (1.25, 0.5) (pl) {};
            \node[anchor=south west] at (pl) {\scriptsize $p_l$};
            \draw (pf) -- (p) -- (pl);
            \node[anchor=north] at (.5,0) {R223};
        \end{tikzpicture}
    \end{minipage}
    \begin{minipage}[b]{.19\linewidth}
        \centering
        \small
        \begin{tikzpicture}
            \tikzset{
                dot/.style={draw, fill=white, circle, inner sep=0pt, minimum size=5},
            }
            \useasboundingbox (-0.5,-0.25) rectangle (1.5,1.5);
            \draw[fill = lightgray!20] (0,0) rectangle (1,1);
            \node[anchor=north west] at (0,1) {$R$};
               
            \node[dot] at (0.75, 0.85) (pf) {};
            
            \node[anchor=north] at (pf) {\scriptsize $p_f$};
            \node[dot] at (1.25, 0.85) (p1) {};
            \node[dot] at (1.25, 0.35) (p2) {};
            \node[dot] at (0.75, 0.35) (pl) {};
            \node[anchor=north] at (pl) {\scriptsize $p_l$};
            \node[anchor=east] at (pl) {$T$};
            \draw (pf) -- (p1) -- (p2) -- (pl);
            \node[anchor=north] at (.5,0) {R247};
        \end{tikzpicture}
    \end{minipage}
    \begin{minipage}[b]{.19\linewidth}
        \centering
        \small
        \begin{tikzpicture}
            \tikzset{
                dot/.style={draw, fill=white, circle, inner sep=0pt, minimum size=5},
            }
            \useasboundingbox (-0.5,-0.25) rectangle (1.5,1.5);
            \draw[fill = lightgray!20] (0,0) rectangle (1,1);
            \node[anchor=north west] at (0,1) {$R$};i.e., R031, R179, R223, R247, and R255. Without loss of generality, we consider
               
            \node[dot] at (0.5,0.5) (pf) {};
            
            \node[anchor=north] at (pf) {\scriptsize $p_f$};
            \node[dot] at (1.25, 0.5) (pl) {};
            \node[anchor=north] at (pl) {\scriptsize $p_l$};
            \node[anchor=west] at (pl) {$T$};
            \draw (pf) -- (pl);
            \node[anchor=north] at (.5,0) {R255};
        \end{tikzpicture}
    \end{minipage}
    
    \begin{minipage}[b]{.19\linewidth}
        \centering
        \scriptsize
        $\begin{aligned}[t]
            \forall p \in T : p \not\sqsubset R
        \end{aligned}$\\
    \end{minipage}
    \begin{minipage}[b]{.19\linewidth}
        \centering
        \scriptsize
        $\begin{aligned}[t]
            \forall p \in T: p \sqsubset R
        \end{aligned}$
    \end{minipage}
    \begin{minipage}[b]{.19\linewidth}
        \centering
        \scriptsize
        $\begin{aligned}[t]
                & p_f, p_l \not \sqsubset R\\[-2.5pt]
        \land  & \exists p \in T_{\widehat{fl}}: p \sqsubset R
        \end{aligned}$\\
    \end{minipage}
    \begin{minipage}[b]{.19\linewidth}
        \centering
        \scriptsize
        $\begin{aligned}[t]
                & p_f, p_l \sqsubset R\\[-2.5pt]
        \land  & \exists p \in T_{\widehat{fl}}: p \not \sqsubset R
        \end{aligned}$\\
    \end{minipage}
    \begin{minipage}[b]{.19\linewidth}
        \centering
        \scriptsize
        $\begin{aligned}[t]
                & p_f \sqsubset R\\[-2.5pt]
        \land   & p_l \not \sqsubset R
        \end{aligned}$\\
    \end{minipage}
    \caption{Visualizations of DE-9IM relationships R031, R179, R223, R247, and R255 with their equivalent predicates.}
    \label{fig:spatial-completeness}
\end{figure}

Regarding spatial relations, we consider the \emph{Dimensionally Extended Nine-Intersection Model} (DE-9IM)~\cite{egenhofer1990mathematical}, a widely utilized topological model to define and reason about spatial relationships between geometric shapes.
The model defines nine intersection patterns regarding the \emph{interior}, \emph{boundary}, and \emph{exterior} between two geometric objects in two dimensions to characterize their spatial relation.
Note that in case one of the geometric objects as a linestring, e.g., the spatial component of a trajectory, the interior is the linestring itself, and the boundary is empty.
While the DE-9IM consists of 6,561 distinct relations between pairs of shapes, Zlatanova et~al.~\cite{zlatanova2016topological}
demonstrate that only 19 relationship types are necessary to model all possible relationships between polygons and linestrings.
Since in our problem setting spatial range queries are relationships between polygons and linestrings, our predicate logic can be considered complete because it can express the 19 relations mentioned above.
Figure~\ref{fig:spatial-completeness} demonstrates, with five examples, how DE-9IM relationships can be expressed as predicates.

\begin{figure}[t]
    \begin{minipage}[b]{.32\linewidth}
        \centering
        \scriptsize
        \begin{tikzpicture}
            \draw[very thick] (-1, .5) -- (-0.25, .5) node[midway, above] {T};
            \draw[very thick] (0.25, 0) -- (1,0) node[midway, above] {I};
            \node[anchor=north] at (0,-0.1) {T precedes I};
        \end{tikzpicture}
    \end{minipage}
    \begin{minipage}[b]{.32\linewidth}
        \centering
        \scriptsize
        \begin{tikzpicture}
            \draw[very thick] (-1, .5) -- (0.25, .5) node[midway, above] {T};
            \draw[very thick] (-0.25, 0) -- (1,0) node[midway, above] {I};
            \node[anchor=north] at (0,-0.1) {T overlaps with I};
        \end{tikzpicture}
    \end{minipage}
    \begin{minipage}[b]{.32\linewidth}
        \centering
        \scriptsize
        \begin{tikzpicture}
            \draw[very thick] (-0.5, .5) -- (0.5, .5) node[midway, above] {T};
            \draw[very thick] (-1, 0) -- (1,0) node[midway, above] {I};
            \node[anchor=north] at (0,-0.1) {T is during I};
        \end{tikzpicture}
    \end{minipage}
    
    \begin{minipage}[b]{.32\linewidth}
        \centering
        \scriptsize
        $\begin{aligned}[t]
            \;\;    & \; \forall p \in T: p < I
        \end{aligned}$\\
    \end{minipage}
    \begin{minipage}[b]{.32\linewidth}
        \centering
        \scriptsize
        $\begin{aligned}[t]
            p_l \sqsubset I \land \exists p \in T_{\widehat{fl}}: p < I
        \end{aligned}$\\
    \end{minipage}
    \begin{minipage}[b]{.32\linewidth}
        \centering
        \scriptsize
        $\begin{aligned}[t]
            \forall p \in T: p \sqsubset I
        \end{aligned}$\\
    \end{minipage}
    \caption{Visualizations of time intervals ``precedes'', ``overlaps with'', and ``is during'' with their equivalent predicates.}
    \label{fig:temporal-completeness}
\end{figure}
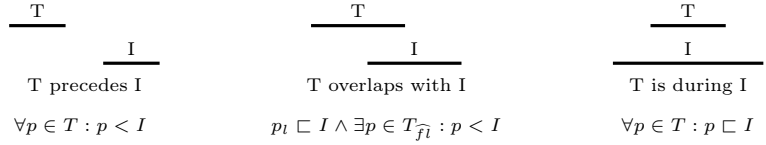

Regarding temporal relations, we focus on \emph{Allen's Interval Algebra}~\cite{allen1983maintaining}, which provides a formal framework for representing and reasoning about temporal intervals. The algebra defines thirteen possible binary relations between time intervals.
With our temporal predicates it is possible to express all interval relations defined in Allen's Interval algebra.
Figure~\ref{fig:temporal-completeness} shows three examples of interval relations expressed as predicates.
A complete list of Allen's Interval Algebra, and DE-9IM relationships along with the corresponding predicates, can be found in Tables~\ref{tab:predicate-allen} and \ref{tab:de9im-relationships}.

As a proof of concept, we consider the spatio-temporal range query and we show how we can express the spatial and temporal predicates in the NF\textsuperscript{2} algebra.

\begin{definition}[Spatio-temporal Range Query]
Given a trajectories relation $\mathfrak{T}(\underline{tid},T(\underline{order},x,y,\tau))$
a spatial region $R$, a time interval $I$, a spatial predicate $P_s$, and a temporal predicate $P_\tau$, a spatio-temporal range query returns all tuples $tp \in \mathfrak{T}$ such that the sequence of $x{\text{-}}y$ coordinates of $tp.T$ and $R$ satisfy $P_s$, and the sequence of timestamps $\tau$ of $tp.T$ and $I$ satisfy $P_\tau$.
\end{definition}
For instance, the relation containing the set of $x{\text{-}}y$ coordinates of a trajectory with $tid=1$ is
\[
\pi[x,y](\mu_{T}(\sigma[tid{=}1](\pi[tid,\pi[x,y](T)](\mathfrak{T})))).
\]

\subsubsection{Spatial Trajectory Selection in NF\textsuperscript{2}}

We begin by showing how NF\textsuperscript{2} algebra can be used to answer queries involving spatial predicates. Note that relationships which involve points or lines lying on the border of a query region are not very useful in practice.
Hence, due to the limited space, we focus on the five relationships that do not involve points or lines on the border of the query region, i.e., R031, R179, R223, R247, and R255\footnote{The DE-9IM relation numbers are based on Zlatanova et~al.~\cite{zlatanova2016topological}}. Without loss of generality, we consider the query region to be a rectangle $R = (x_{min},y_{min}, x_{max}, y_{max})$. 

We begin with relations R179, R247, and R255. For these relations, the algebraic expressions for both strict and relaxed evaluation are the same. Wherever necessary in the following examples we have:
\begin{align*}
P_{first} = & \sigma[order = 1](T) \\
P_{last} = & \sigma[order = \max(\pi[order](T'))](T)
\end{align*}

\begin{example}
Given a trajectories relation $\mathfrak{T}$ and a query rectangle $R$, the relationship R179 returns all trajectories/tuples $tp \in \mathfrak{T}$ such that the trajectory $T$ lies completely inside $R$. The NF\textsuperscript{2} algebra expression for this query is
\begin{align*}
\sigma^{ST}_{R179, R} = & \sigma[
x_{min}{<}\min(\pi[x](T)) \land 
y_{min}{<}\min(\pi[y](T)) \land \\
& \;\;\; x_{max}{>}\max(\pi[x](T)) \land
y_{max}{>}\max(\pi[y](T)) \; ](\mathfrak{T})
\end{align*}
\end{example} 

\begin{example}
Given a trajectories relation $\mathfrak{T}$ and a query rectangle $R$, the relationship R247 returns all trajectories $tp \in \mathfrak{T}$ the starting point $p_f$ and the ending point $p_l$ lie completely inside $R$ and there exists at least one point $p$ of $t$ that lies outside $R$. The NF\textsuperscript{2} algebra expression for this query is
\begin{align*}
\sigma^{ST}_{R247, R} =
& \sigma[\pi[x](P_{first}){>}x_{min} \land \pi[x](P_{first}){<}x_{max} \land\\
& \;\; \pi[y](P_{first}){>}y_{min} \land \pi[y](P_{first}){<}y_{max} \land \\
& \;\; \pi[x](P_{last}){>}x_{min} \land \pi[x](P_{last}){<}x_{max} \land\\
& \;\; \pi[y](P_{last}){>}y_{min} \land \pi[y](P_{last}){<}y_{max} \land (\\
& \;\;\;\;\;\; \min(\pi[x](T)){<}x_{min} \lor \max(\pi[x](T)){>}x_{max} \lor\\
& \;\;\;\;\;\; \min(\pi[y](T)){<}y_{min} \lor \max(\pi[y](T)){>}y_{max}) \; ](\mathfrak{T})
\end{align*}
\end{example} 

\begin{example}
Given a trajectories relation $\mathfrak{T}$ and a query rectangle $R$, the relationship R255 returns all trajectories $tp \in \mathfrak{T}$ the starting point $p_f$ of which lies inside $R$ and the ending point $p_l$ of which lies outside $R$. The NF\textsuperscript{2} algebra expression for this query is
\begin{align*}
\sigma^{ST}_{R255, R} = 
& \sigma[\pi[x](P_{first}){>}x_{min} \land \pi[x](P_{first}){<}x_{max} \land\\
& \;\; \pi[y](P_{first}){>}y_{min} \land \pi[y](P_{first}){<}y_{max} \land (\\
& \;\;\;\;\;\; \min(\pi[x](P_{last})){<}x_{min} \lor \max(\pi[x](P_{last})){>}x_{max} \lor\\
& \;\;\;\;\;\; \min(\pi[y](P_{last})){<}y_{min} \lor \max(\pi[y](P_{last})){>}y_{max}) \; ](\mathfrak{T})
\end{align*}
\end{example} 

We now focus on relations R031, and R223. As shown in Figure~\ref{fig:spatial-completeness}, in order to apply the relaxed evaluation of these predicates, one must check every line segment form by consecutive trajectory points. In order to enable the examination of trajectory line segments, we define the following relation that we use in the subsequence examples:
\begin{align*}
T_{sgmt} = & \pi[T.order, T.x, T.y, T'.x, T'.y] (T \bowtie_{T.order+1 = T'.order} T')
\end{align*}

\begin{example}
Given a trajectories relation $\mathfrak{T}$ and a query rectangle $R$, the relationship R031 returns all trajectories $tp \in \mathfrak{T}$ that lie completely outside $R$. For the strict evaluation, it is sufficient to check that all points of each $T$ lie outside $R$. As such, the NF\textsuperscript{2} algebra expression for this query is
\begin{align*}
\sigma^{ST}_{R031, R} = & \sigma[\text{count}(\sigma[x_{min}{<}x \land y_{min}{<}y \land x_{max}{>}x \land y_{max}{>}y](T)) = 0 \; ](\mathfrak{T})
\end{align*}
\end{example} 
For the relaxed evaluation, we also need to check that none of the segments of $T$ in $T_{sgmt}$ intersect $R$.

\begin{example}
Given a trajectories relation $\mathfrak{T}$ and a query rectangle $R$, the relationship R223 returns all trajectories $tp \in \mathfrak{T}$ the starting point $p_f$ and the ending point $p_l$ of the associated trajectory $T$ that lie outside $R$ and $T$ intersects $R$. For the strict evaluation, it is sufficient to check that $p_f$ and $p_l$ lie outside $R$ and there is at least one point of $T$ that lies inside $R$. As such, the NF\textsuperscript{2} algebra expression for this query is
\begin{align*}
\sigma^{ST}_{R223, R} = & \sigma[( \; \pi[x](P_{first}){<}x_{min} \lor \pi[y](P_{first}){<}y_{min} \lor \\
& \;\;\;\;\;\; \pi[x](P_{first}){>}x_{max} \lor \pi[y](P_{first}){>}y_{max}) \land \\
& \;\;\;\; ( \, \pi[x](P_{last}){<}x_{min} \lor \pi[y](P_{last}){<}y_{min} \lor \\
& \;\;\;\;\;\;\pi[x](P_{last}){>}x_{max} \lor \pi[y](P_{last}){>}y_{max}) \land \\
& \;\; \text{count}(\sigma[x{>}x_{min} \land y{>}y_{min} \land x{<}x_{max} \land y{<}y_{max}](T)){>}0 \; ](\mathfrak{T})
\end{align*}
\end{example} 
For the relaxed evaluation, in the case where $\text{count} = 0$ we need to check whether at least one of the segments in $T_{sgmt}$ intersects $R$.

\subsubsection{Temporal Trajectory Selection in NF\textsuperscript{2}}

We now show how the NF\textsuperscript{2} algebra can be used to answer queries involving temporal predicates. Due to the limited space, similar to the spatial trajectory selection, we focus on relationships that do not consider points lying at the start or the end of a given interval.
Hence, we focus on the relationships \emph{precedes}, \emph{overlaps with}, and \emph{is during}. Without loss of generality, we consider the query interval $I = (\tau_{s},\tau_{e})$.

\begin{example}
Given a trajectories relation $\mathfrak{T}$ and a query interval $I$, the \emph{precedes} relationship returns all trajectories $tp \in \mathfrak{T}$ with a last point that has a timestamp before the starting timestamp $\tau_{s}$ of the interval $I$. The NF\textsuperscript{2} algebra expression for this query is
\begin{align*}
\sigma^{ST}_{\emph{precedes}, I} = & \sigma[ \max(\pi[\tau](T)){<}\tau_{s}](\mathfrak{T}).
\end{align*}
\end{example}

\begin{example}
Given a trajectories relation $\mathfrak{T}$ and a query interval $I$, the \emph{overlaps with} relationship returns all trajectories $tp \in \mathfrak{T}$ for which the timestamp of the first point is before the start $\tau_s$ of the interval $I$, and the timestamp of the last point is after $\tau_{s}$ but before the end $\tau_e$ of the interval $I$. The NF\textsuperscript{2} algebra expression for this query is
\begin{align*}
\sigma^{ST}_{\emph{overlaps with}, I} = & \sigma[ \min(\pi[\tau](T)){<}\tau_{s} \land \max(\pi[\tau](T)){>}\tau_{s} \land \min(\pi[\tau](T)){<}\tau_{e}](\mathfrak{T}).
\end{align*}
\end{example} 

\begin{example}
Given a trajectories relation $\mathfrak{T}$ and a query interval $I$, the \emph{is during} relationship returns all trajectories $tp \in \mathfrak{T}$ for which both the timestamp of the first point and the timestamp of the last point are after the start $\tau_s$ of the interval $I$ and before the end $\tau_e$ of $I$. The NF\textsuperscript{2} algebra expression for this query is
\begin{align*}
\sigma^{ST}_{\emph{overlaps with}, I} = & \sigma[ \min(\pi[\tau](T)){>}\tau_{s} \land \max(\pi[\tau](T)){<}\tau_{e} ](\mathfrak{T}).
\end{align*}
\end{example} 

In a similar fashion, we can define the mirrored relationships \emph{is preceded by}, \emph{is overlapped by}, and \emph{contains}.
Note that for the temporal dimension, there is no difference between the strict and the relaxed evaluation since we only need to consider the start and the end timestamp of a trajectory.

\section{Conclusion}\label{sec:conclusion}
This paper proposes a formal data model and predicate logic for unified trajectory data management.
Introducing a novel data model rooted in the NF\textsuperscript{2} relational data model, we merge spatial, temporal, and application-specific attributes of trajectories.
Our unified spatio-temporal predicate logic handles uncertainty in sampled trajectory data by accommodating varying levels of strictness.

Regarding future work, this paper lays the formal foundation for a query broker for trajectory data.
Additionally, we aim to design a specialized query optimization framework to improve the broker's efficiency and scalability in handling complex trajectory queries, paving the way for enhanced trajectory data management and analysis capabilities.

\appendix

\section*{Acknowledgment}
This work is funded by the Deutsche Forschungsgemeinschaft (DFG) under Germany's Excellence Strategy -- EXC 2117 -- 422037984.

\bibliographystyle{splncs04}
\bibliography{related}

\section*{Appendix}
Table~\ref{tab:predicate-allen} illustrates all relationships defined in Allen's interval algebra, while Table~\ref{tab:de9im-relationships} illustrates the full list of DE-9IM line-to-area relationships.

\begin{table}
    \caption{Allen's interval algebra and expressions using the operators of Section~\ref{sec:temporal-predicates}.}
    \label{tab:predicate-allen}
    \tikzset{
        box/.style={fill=lightgray!20},
        dot/.style={draw, circle, fill,inner sep=0pt, minimum size=5},
    }
    \centering
    \renewcommand{\arraystretch}{1.6}
    \begin{tabular}{ | c | c | l | }
        \hline
        \multirow{2}{*}{
        \begin{tikzpicture}
        \draw[very thick] (-1, .5) -- (-0.25, .5) node[midway, above] {T}; \draw[very thick] (0.25, 0) -- (1,0) node[midway, above] {I}; 
        \end{tikzpicture}
        } 
        & T precedes I &
        \multirow{2}{*}{
        $\begin{aligned} 
        &  \forall p \in T: p < I
        \end{aligned}$} \\
        & \;\; I is preceded by T \;\; &  \\
        
        \hline
        
        \multirow{2}{*}{
        \begin{tikzpicture}
        \draw[very thick] (-1, .5) -- (0, .5) node[midway, above] {T}; \draw[very thick] (0, 0) -- (1,0) node[midway, above] {I};
        \end{tikzpicture}
        } 
        & T meets I &
        \multirow{2}{*}{
        $\begin{aligned} 
            & p_l \sqsubseteq I \land \neg ( p_l \sqsubset I ) \land  \forall p \in T_{\widehat{fl}}: p < I
            \end{aligned}$} \\
        & I is met by T &  \\
        
        \hline
        
        \multirow{2}{*}{
        \begin{tikzpicture}
        \draw[very thick] (-1, .5) -- (0.25, .5) node[midway, above] {T}; \draw[very thick] (-0.25, 0) -- (1,0) node[midway, above] {I};
        \end{tikzpicture}
        } 
        & T overlaps with I &
        \multirow{2}{*}{
        $\begin{aligned} 
                & p_l \sqsubset I \land \exists p \in T_{\widehat{fl}}: p < I
        \end{aligned}$} \\
        & I is overlapped by T &  \\

        \hline
        
        \multirow{2}{*}{
        \begin{tikzpicture}
        \draw[very thick] (-1, .5) -- (0, .5) node[midway, above] {T}; \draw[very thick] (-1, 0) -- (1,0) node[midway, above] {I};
        \end{tikzpicture}
        } 
        & T starts I &
        \multirow{2}{*}{
        $\begin{aligned} 
                & p_f \sqsubseteq I
        \land \neg ( p_f \sqsubset I )
        \land p_l \sqsubset I
        \end{aligned}$} \\
        & I is started by T &  \\

        \hline
        
        \multirow{2}{*}{
        \begin{tikzpicture}
        \draw[very thick] (-0.5, .5) -- (0.5, .5) node[midway, above] {T}; \draw[very thick] (-1, 0) -- (1,0) node[midway, above] {I};
        \end{tikzpicture}
        } 
        & T is during I &
        \multirow{2}{*}{
        $\begin{aligned} 
                &  \forall p \in T: p \sqsubset I
        \end{aligned}$} \\
        & I contains T &  \\

        \hline
        
        \multirow{2}{*}{
        \begin{tikzpicture}
        \draw[very thick] (0, .5) -- (1, .5) node[midway, above] {T}; \draw[very thick] (-1, 0) -- (1,0) node[midway, above] {I};
        \end{tikzpicture}
        } 
        & T finishes I &
        \multirow{2}{*}{
        $\begin{aligned} 
                & p_l \sqsubseteq I
        \land   \neg ( p_l \sqsubset I )
        \land  p_f \sqsubset I
        \end{aligned}$} \\
        & I is finished by T &  \\

        \hline
        
        \multirow{2}{*}{
        \begin{tikzpicture}
        \draw[very thick] (-1, .5) -- (1, .5) node[midway, above] {T}; \draw[very thick] (-1, 0) -- (1,0) node[midway, above] {I};
        \end{tikzpicture}
        } 
        & \multirow{2}{*}{T is equal to I} &
        \multirow{2}{*}{
        $\begin{aligned} 
                &  p_f \sqsubseteq I 
        \land   \neg ( p_f \sqsubset I )
        \land   p_l \sqsubseteq I 
        \land   \neg ( p_l \sqsubset I ) 
        \end{aligned}$} \\
        & &  \\
        \hline
        
    \end{tabular}
    
\end{table}

\begin{table}
    \caption{DE-9IM Relationships and expressions using the operators of Section~\ref{sec:spatial-predicates}.}
    \label{tab:de9im-relationships}
    \tikzset{
        box/.style={fill=lightgray!20},
        dot/.style={draw, circle, fill,inner sep=0pt, minimum size=5},
    }
    \small
    \centering
    \begin{tabular}{ | m{1.1cm} | c | l || m{1.1cm} | c | l |}
        \hline
        \scalebox{.5}{\tikz{\node at (-.5,-.5) {}; \node at (1.5,1.5) {}; \draw [box] (0,0) rectangle (1,1); \node[dot] at (-.5, 1) (a) {}; \node[dot] at (0, 1.5) (b) {}; \draw[very thick] (a) -- (b);}}
        &
        \; R031 \; 
        &
        $\begin{aligned} 
            \;\;    & \; \forall p \in T : p \not\sqsubset R
        \end{aligned}$
        &
        \scalebox{.5}{\tikz{\node at (-.5,-.5) {}; \node at (1.5,1.5) {}; \draw [box] (0,0) rectangle (1,1); \node[dot] at (0.5, 0) (a) {}; \node[dot] at (1, 0.5) (b) {}; \draw[very thick] (a) -- (0.5, -0.25) -- (1, -0.25) --(b);}}
        & 
        \; R343 \; 
        &
        $\begin{aligned} 
        & p_f, p_l \sqsubseteq R \land \neg ( p_f, p_l \sqsubset R ) \land\\[-2.5pt]
        & \hspace{-4pt} \not\exists p \in T_{\widehat{fl}}: p \sqsubset R \land \\[-2.5pt]
        & \exists p_1 \in T_{\widehat{fl}}: p_1 \sqsubseteq R \land \\[-2.5pt]
        & \exists p_2 \in T_{\widehat{fl}}: p_2 \not \sqsubset R
        \end{aligned}$
        \\ \hline
        \scalebox{.5}{\tikz{\node at (-.5,-.5) {}; \node at (1.5,1.5) {}; \draw [box] (0,0) rectangle (1,1); \node[dot] at (0.25, 1.25) (a) {}; \node[dot] at (0.75, 1.25) (b) {}; \draw[very thick] (a) -- (.33, 1) -- (.66, 1) -- (b);}}
        & 
        R095
        &
        $\begin{aligned} 
                & \; p_f, p_l \not \sqsubset R\\[-2.5pt]
        \wedge  & \; \exists p \in T_{\widehat{fl}}: p \sqsubseteq R\\[-2.5pt]
        \wedge  & \; \neg ( p \sqsubset R )
        \end{aligned}$
        &
        \scalebox{.5}{\tikz{\node at (-.5,-.5) {}; \node at (1.5,1.5) {}; \draw [box] (0,0) rectangle (1,1); \node[dot] at (0.5, -0.25) (a) {}; \node[dot] at (1, 1) (b) {}; \draw[very thick] (a) -- (1, 0) -- (b);}}
        & 
        R351
        &
        $\begin{aligned} 
                & \; p_f \sqsubseteq R  \\[-2.5pt]
        \wedge  & \; \neg ( p_f \sqsubset R ) \\[-2.5pt]
        \wedge  & \; p_l \not \sqsubset R  \\[-2.5pt]
        \wedge  & \; \not \exists p \in T_{\widehat{fl}}: p \sqsubset R\\[-2.5pt]
        \wedge  & \; \exists p_1 \in T_{\widehat{fl}}: p_1 \sqsubseteq R \\[-2.5pt]
        \wedge  & \; \exists p_2 \in T_{\widehat{fl}}: p_2 \not \sqsubset R
        \end{aligned}$
        \\ \hline
        \scalebox{.5}{\tikz{\node at (-.5,-.5) {}; \node at (1.5,1.5) {}; \draw [box] (0,0) rectangle (1,1); \node[dot] at (.25, 0.25) (a) {}; \node[dot] at (0.75, 0.75) (b) {}; \draw[very thick] (a) -- (b);}}
        & 
        R179
        &
        $\begin{aligned} 
            \;\;    & \; \forall p \in T: p \sqsubset R
        \end{aligned}$
        &
        \scalebox{.5}{\tikz{\node at (-.5,-.5) {}; \node at (1.5,1.5) {}; \draw [box] (0,0) rectangle (1,1); \node[dot] at (0.5, 0) (a) {}; \node[dot] at (0.5, 1) (b) {}; \draw[very thick] (a) -- (b);}}
        & 
        R403
        &
        $\begin{aligned} 
                & \; p_f, p_l \sqsubseteq R\\[-2.5pt]
        \wedge  & \; \neg (p_f, p_l \sqsubset R) \\[-2.5pt]
        \wedge  & \;  \forall p \in T_{\widehat{fl}} : p \sqsubset R
        \end{aligned}$
        \\ \hline
        \scalebox{.5}{\tikz{\node at (-.5,-.5) {}; \node at (1.5,1.5) {}; \draw [box] (0,0) rectangle (1,1); \node[dot] at (.3, -.25) (a) {}; \node[dot] at (1.25, .7) (b) {}; \draw[very thick] (a) -- (b);}}
        & 
        R223
        &
        $\begin{aligned} 
                & \; p_f, p_l \not \sqsubset R\\[-2.5pt]
        \wedge  & \; \exists p \in T_{\widehat{fl}}: p \sqsubset R
        \end{aligned}$
        &
        \scalebox{.5}{\tikz{\node at (-.5,-.5) {}; \node at (1.5,1.5) {}; \draw [box] (0,0) rectangle (1,1); \node[dot] at (0.5, 0.5) (a) {}; \node[dot] at (0.5, 1) (b) {}; \draw[very thick] (a) -- (b);}}
        & 
        R435
        &
        $\begin{aligned} 
                & \; p_f \sqsubseteq R\\[-2.5pt]
        \wedge  & \; \neg ( p_f \sqsubset R )\\[-2.5pt]
        \wedge  & \; p_l \sqsubset R\\[-2.5pt]
        \wedge  & \; \forall p \in T_{\widehat{fl}}: p \sqsubset R
        \end{aligned}$
        \\ \hline
        \scalebox{.5}{\tikz{\node at (-.5,-.5) {}; \node at (1.5,1.5) {}; \draw [box] (0,0) rectangle (1,1); \node[dot] at (.25, .3) (a) {}; \node[dot] at (.75, .6) (b) {}; \draw[very thick] (a) -- (.25, 1) -- (.75, 1) -- (b);}}
        & 
        R243
        &
        $\begin{aligned} 
                & \; p_f, p_l \sqsubset R \\[-2.5pt]
        \wedge  & \; \forall p \in T_{\widehat{fl}}: p \sqsubseteq R \\[-2.5pt]
        \wedge  & \; \exists p \in T_{\widehat{fl}}: \neg( p \sqsubset R )
        \end{aligned}$
        &
        \scalebox{.5}{\tikz{\node at (-.5,-.5) {}; \node at (1.5,1.5) {}; \draw [box] (0,0) rectangle (1,1); \node[dot] at (0.5, 0.25) (a) {}; \node[dot] at (1, 0.75) (b) {}; \draw[very thick] (a) -- (1, 0.25) -- (b);}}
        & 
        R467
        &
        $\begin{aligned} 
                & \; p_f \sqsubseteq R\\[-2.5pt]
        \wedge  & \; \neg ( p_f \sqsubset R ) \\[-2.5pt]
        \wedge  & \; p_l \sqsubset R\\[-2.5pt]
        \wedge  & \; \forall p \in T_{\widehat{fl}}: p \sqsubseteq R\\[-2.5pt]
        \wedge  & \; \exists p_1 \in T_{\widehat{fl}}: p_1 \sqsubset R\\[-2.5pt]
        \wedge  & \; \exists p_2 \in T_{\widehat{fl}}: \neg (p_2 \sqsubset R)
        \end{aligned}$
        \\ \hline
        \scalebox{.5}{\tikz{\node at (-.5,-.5) {}; \node at (1.5,1.5) {}; \draw [box] (0,0) rectangle (1,1); \node[dot] at (0.25, 0.3) (a) {}; \node[dot] at (0.75, 0.6) (b) {}; \draw[very thick] (a) -- (.25, 1.25) -- (.75, 1.25) -- (b);}}
        & 
        R247
        &
        $\begin{aligned} 
                & \; p_f, p_l \sqsubset R\\[-2.5pt]
        \wedge  & \; \exists p \in T_{\widehat{fl}}: p \not \sqsubset R
        \end{aligned}$
        &
        \scalebox{.5}{\tikz{\node at (-.5,-.5) {}; \node at (1.5,1.5) {}; \draw [box] (0,0) rectangle (1,1); \node[dot] at (0.25, 1) (a) {}; \node[dot] at (0.75, 0) (b) {}; \draw[very thick] (a) -- (0.25, 1.25) -- (0.75, 1.25) -- (b);}}
        & 
        R471
        &
        $\begin{aligned} 
                & \; p_f, p_l \sqsubseteq R  \\[-2.5pt]
        \wedge  & \; \neg ( p_f, p_l \sqsubset R )  \\[-2.5pt]
        \wedge  & \;\exists p_1 \in T_{\widehat{fl}}: p_1 \sqsubset R \\[-2.5pt]
        \wedge  & \; \exists p_2 \in T_{\widehat{fl}}: p_2 \not \sqsubset R 
        \end{aligned}$
        \\ \hline
        \scalebox{.5}{\tikz{\node at (-.5,-.5) {}; \node at (1.5,1.5) {}; \draw [box] (0,0) rectangle (1,1); \node[dot] at (0.5, 0.5) (a) {}; \node[dot] at (0.5, 1.25) (b) {}; \draw[very thick] (a) -- (b);}}
        & 
        R255
        &
        $\begin{aligned} 
                & \; p_f \sqsubset R\\[-2.5pt]
        \wedge  & \; p_l \not \sqsubset R
        \end{aligned}$
        &
        \scalebox{.5}{\tikz{\node at (-.5,-.5) {}; \node at (1.5,1.5) {}; \draw [box] (0,0) rectangle (1,1); \node[dot] at (0.5, 1) (a) {}; \node[dot] at (0.5, -0.25) (b) {}; \draw[very thick] (a) -- (b);}}
        & 
        R479
        &
        $\begin{aligned} 
                & \; p_f \sqsubseteq R\\[-2.5pt]
        \wedge  & \; \neg ( p_f \sqsubset R ) \\[-2.5pt]
        \wedge  & \; p_l \not \sqsubset R \\[-2.5pt]
        \wedge  & \; \exists p_1 \in T_{\widehat{fl}}: p_1 \sqsubset R \\[-2.5pt]
        \wedge  & \; \exists p_2 \in T_{\widehat{fl}}: p_2 \not \sqsubset R
        \end{aligned}$
        \\ \hline
        \scalebox{.5}{\tikz{\node at (-.5,-.5) {}; \node at (1.5,1.5) {}; \draw [box] (0,0) rectangle (1,1); \node[dot] at (0.25, 1) (a) {}; \node[dot] at (0.75, 1) (b) {}; \draw[very thick] (a) -- (0.25, 1.25) -- (0.75, 1.25) -- (b);}}
        & 
        R279
        &
        $\begin{aligned} 
                & \; p_f, p_l \sqsubseteq R\\[-2.5pt]
        \wedge  & \; \neg ( p_f, p_l \sqsubset R ) \\[-2.5pt]
        \wedge  & \; \forall p \in T_{\widehat{fl}}: p \not \sqsubset R
        \end{aligned}$
        &
        \scalebox{.5}{\tikz{\node at (-.5,-.5) {}; \node at (1.5,1.5) {}; \draw [box] (0,0) rectangle (1,1); \node[dot] at (0.25, 1) (a) {}; \node[dot] at (0.75, 0) (b) {}; \draw[very thick] (a) -- (0.75, 1) --(b);}}
        & 
        R499
        &
        $\begin{aligned} 
                & \; p_f, p_l \sqsubseteq R\\[-2.5pt]
        \wedge  & \; \neg ( p_f, p_l \sqsubset R )\\[-2.5pt]
        \wedge  & \; \forall p \in T_{\widehat{fl}}: p \sqsubseteq R\\[-2.5pt]
        \wedge  & \; \exists p_1 \in T_{\widehat{fl}}: p_1 \sqsubset R\\[-2.5pt]
        \wedge  & \; \exists p_2 \in T_{\widehat{fl}}: \neg (p_2 \sqsubset R)
        \end{aligned}$
        \\ \hline
        \scalebox{.5}{\tikz{\node at (-.5,-.5) {}; \node at (1.5,1.5) {}; \draw [box] (0,0) rectangle (1,1); \node[dot] at (1, .25) (a) {}; \node[dot] at (1.25, .75) (b) {}; \draw[very thick] (a) -- (b);}}
        & 
        R287
        &
        $\begin{aligned} 
                & \; p_f \sqsubseteq R\\[-2.5pt]
        \wedge  & \; \neg ( p_f \sqsubset R )\\[-2.5pt]
        \wedge  & \; p_l \not \sqsubset R\\[-2.5pt]
        \wedge  & \; \forall p \in T_{\widehat{fl}}: p  \not \sqsubset R
        \end{aligned}$
        &
        \scalebox{.5}{\tikz{\node at (-.5,-.5) {}; \node at (1.5,1.5) {}; \draw [box] (0,0) rectangle (1,1); \node[dot] at (0.25, 1) (a) {}; \node[dot] at (0.75, 0.5) (b) {}; \draw[very thick] (a) -- (0.25, 1.25) -- (0.75, 1.25) -- (b);}} 
        & 
        R503
        &
        $\begin{aligned} 
                & \; p_f \sqsubseteq R\\[-2.5pt]
        \wedge  & \; \neg ( p_f \sqsubset R )  \\[-2.5pt]
        \wedge  & \; p_l \sqsubset R  \\[-2.5pt]
        \wedge  & \; \exists p \in T_{\widehat{fl}}: p \not \sqsubset R
        \end{aligned}$
        \\ \hline
        \scalebox{.5}{\tikz{\node at (-.5,-.5) {}; \node at (1.5,1.5) {}; \draw [box] (0,0) rectangle (1,1); \node[dot] at (.25, 1) (a) {}; \node[dot] at (1, .5) (b) {}; \draw[very thick] (a) -- (1, 1) -- (b);}}
        & 
        R339
        &
        $\begin{aligned} 
                & \; \forall p \in T: p \sqsubseteq R\\[-2.5pt]
        \wedge  & \; \neg( p \sqsubset R )
        \end{aligned}$
        &
        \multicolumn{3}{c|}{}
        \\ \hline
    \end{tabular}
    
\end{table}

\end{document}